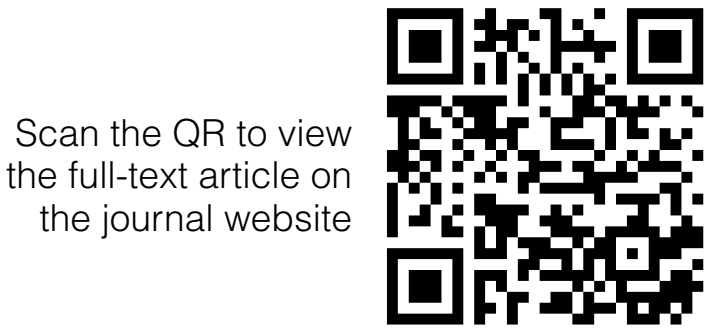

REVIEW

# Comparison of Liu-Type Estimator for Multicollinearity in Fuzzy Logistic Regression Model

**Ayad Habib Shemail** [a,*], **Ahmed Razzaq Al-Lami** [a], **Amal Hadi Rashid** [b]

[a] Department of Statistics, University of Wasit, Wasit, Iraq
[b] Department of Statistics, University of Diyala, Diyala, Iraq

**ABSTRACT**

This article addresses the fuzzy logistic regression model under conditions of multicollinearity, which causes instability and inflated variance in parameter estimation. In this model, both the response variable and parameters are represented as fuzzy triangular numbers. To overcome the multicollinearity problem, various Liu-type estimators were employed: Fuzzy Maximum Likelihood Estimators (FMLE), Fuzzy Logistic Ridge Estimators (FLRE), Fuzzy Logistic Liu Estimators (FLLE), Fuzzy Logistic Liu-type Estimators (FLLTE), and Fuzzy Logistic Liu-type Parameter Estimators (FLLTPE). Through simulations with various sample sizes and application to real fuzzy data on kidney failure, model performance was evaluated using mean square error (MSE) and goodness of fit criteria. Results demonstrated superior performance of FLLTPE and FLLTE compared to other estimators.

**Keywords:** Fuzzy logistic regression model, Multicollinearity problem, Fuzzy triangular number, Fuzzy maximum likelihood estimators, Fuzzy logistic Liu-type estimators

## 1. Introduction

Collinearity between explanatory variables is a major challenge in fuzzy logistic regression analysis. Especially when estimating parameters in models with highly correlated predictors. This phenomenon amplifies the variance of parameter estimates. Leading to unreliable and unstable results. Dealing with multicollinearity is crucial to obtaining accurate and specific estimates in fuzzy regression models. Especially in fields where predictive performance is critical, Such as economics, Science, and medicine.

In response to the issue of multicollinearity, various estimation techniques have been proposed and evaluated over the years. Among these techniques, Liu-type estimators have emerged as a promising approach to mitigate the negative effects of multicollinearity and improve the predictive performance of regression models. Liu-type estimators work to reduce the error of the parameter estimates in a regression model towards certain target values. Thereby reducing the variance of the estimates and enhancing their stability.

The effectiveness of Liu-type estimators has been demonstrated across various regression models. Including linear regression. Poisson. Logistic and fuzzy models. These estimators have shown superiority over traditional estimators, such as maximum likelihood and ridge estimators. Under multicollinearity conditions. The researchers provided evidence of the superior performance of Liu-type estimators in terms of accuracy Precision.

The Liu-type estimator, proposed by [11], demonstrates a smaller mean square error compared to the maximum likelihood estimator in logistic regression models. the Liu-type estimator and several alternative estimators, namely, the letter estimators and the principal components estimators in logistic regression. are compared under the mean square error criterion if multicollinearity among the independent variables is found to inflate the variance, and the Liu-type estimator outperforms the other estimators. It is known that multicollinearity inflates the variance of the maximum likelihood estimators in logistic regression, especially. suppose the primary interest is in the coefficients. the effect of multicollinearity can be very serious.

* Corresponding author.
E-mail addresses: ashemail@uowasit.edu.iq (A. H. Shemail), ahmedrazzaq@uowasit.edu.iq (A. R. Al-Lami), amal@uodiyala.edu.iq (A. H. Rashid).

Researchers [3], presented a Liu-type estimator that had a smaller mean square error (MSE) than the latter estimator under certain conditions. Simulation studies were conducted to evaluate the performance of this estimator, and the proposed estimator was applied to a real-life dataset. Researchers [17], introduced new shrinkage parameters for Liu-type estimators in the logistic regression model defined by [8], in order to reduce the variance and overcome the multicollinearity problem among explanatory variables. A simulation study is designed to evaluate the quality of the proposed estimators over the maximum likelihood estimator (MLE) by the mean absolute square error (AMSE). In addition. A real data case is presented to demonstrate the advantages of the new shrinkage parameters.

Researchers [12], have proposed several alternative estimators in the literature when some linear constraints on the parameters are available to overcome the multicollinearity problem in logistic regression, a new two-parameter Liu-type estimator called the stochastic restricted logistic Liu estimator (SRLTLE) is proposed by combining Liu estimator and logistic model in the presence of stochastic linear constraints. Moreover, a Monte Carlo simulation study is conducted to compare the performance of the proposed estimator with some classical estimators through the standard mean square error (SMSE); a numerical example is presented to illustrate the theoretical results.

Researchers [10], proposed a Liu-type estimator in the logistic regression considered a general estimator that includes two estimators: the Liu estimator and the ridge estimator. The proposed estimator is characterized by more efficiency and reliability, in addition to reducing variance than classical estimators.

Researchers [4], presented a new biased estimator to combat multicollinearity in the logistic regression model. The proposed estimator is a general estimator that includes two other biased estimators. Logistic Ridge and Logistic Liu, with two bias coefficients as special cases. The necessary and sufficient conditions for the new biased estimator to outperform the existing estimators are obtained. Monte Carlo simulation studies are also performed to compare the performance of the proposed biased estimator. Finally, a numerical example is presented to illustrate some theoretical results in the same year.

Researchers [16], developed a Liu-type estimator that combines the ridge estimator and the Liu-type estimator to reduce variance inflation and solve the multicollinearity problem in the logistic regression model. Based on a Monte Carlo simulation study to evaluate the performance of the ridge estimator with the Liu-type estimator, through the MSE criterion and bias criteria, it was concluded that the new estimators perform well when combined with the Liu-type estimator.

Also showing the Homotopy Analysis Method, which has been successfully applied to fuzzy nonlinear Volterra integral equations [13], is one of the sophisticated analytical techniques used in a number of recent studies that address fuzzy systems.

It can be noted from a study [19], under multicollinearity conditions, a novel Liu-type estimator in the context of IGR significantly decreased the mean squared error when compared to classical methods.

Therefore, the purpose of this study is to use Liu-type estimators to estimate the parameters of the fuzzy logistic regression model in the presence of multicollinearity. The study applies these estimators to actual kidney failure data and uses simulation techniques to assess their effectiveness. This study is important because it offers accurate and reliable estimation methods for efficiently evaluating fuzzy data, especially in scientific and medical domains where making accurate decisions is essential.

## 2. Fuzzy logic

Fuzzy logic is a mathematical logic designed to deal with vague and imprecise concepts. unlike traditional logic that relies on binary values (true/false). Zadeh introduced the concept of fuzzy logic in 1965. and it is an extension of traditional logic that relies on probability theory. Fuzzy logic aims to deal with uncertainty and ambiguity in data. i.e, it allows solving problems using imprecise and open sets of data to reach accurate conclusions [7].

One of the forms of fuzzy logic is the fuzzy set. which is defined as a set of numbers z that belong to the universal set Z with a certain membership function $\mu_{\tilde{A}(z)}$ i.e. they are in the form of ordered pairs. so, the fuzzy set $\tilde{A}$ is defined as follows [5].

$$\tilde{A} = \left\{ \left( z . \mu_{\tilde{A}(z)} \right) \backslash z \in Z \right\} \tag{1}$$

Where $\mu_{\tilde{A}(z)}$ represents the degree of belonging of element z to the fuzzy set $\tilde{A}$ and i.e., it belongs to the interval [0,1].

Let $\widetilde{Z} = (l_z.z.r_z)$ is defined with the membership function given in Eq. (2). The graphical representation of this triangular fuzzy number is illustrated in Fig. 1 [1].

$$\mu_{\tilde{Z}(x)} = \begin{cases} 0 & for\ x < l_z \\ \frac{x-l_z}{z-l_z} & for\ l_z \le x \le z \\ \frac{r_z-x}{r_z-z} & for\ z \le x \le r_z \\ 0 & for\ x > r_z \end{cases} \tag{2}$$

The above belonging function has the following conditions:

1. $\mu(z)$ *is an increasing function for z.*
2. $r_z$ is a decreasing function for to z.
3. $l_z \le z \le r_z$

A triangular fuzzy number $\tilde{Z} = (l_z.z.r_z)$ is considered positive if $l_z.z.r_z > 0$ and a negative triangular fuzzy number if $l_z.z.r_z < 0$.

Let $\widetilde{Z} = (l_z.z.r_z)$, $\widetilde{C} = (l_c.c.r_c)$ be two triangular fuzzy numbers. The mathematical operations on fuzzy numbers that we need in the logistic regression model are

1- Equal: $\tilde{Z} = \tilde{C}$ If and only if $l_z = l_c.z = c$ and $r_z = r_c$.

2- Addition:

$$\tilde{Z} + \tilde{C} = (l_z + l_c.z + c.r_z + r_c) \tag{3}$$

3- Subtraction:

$$\tilde{Z} - \tilde{C} = (l_z - l_c.z - c.r_z - r_c) \tag{4}$$

4- Multiplication:

$$\widetilde{Z} * \tilde{C} = (\min(l_z l_c.\ l_z r_c.\ r_z l_c.r_z r_c).zc.\ max(l_z l_c.\ l_z r_c.\ r_z l_c.r_z r_c)) \tag{5}$$

5- Division:

$$\tilde{Z}/_{\tilde{C}} = \left(\min\left(\frac{l_z}{l_c}.\frac{l_z}{r_c}.\frac{r_z}{l_c}.\frac{r_z}{r_c}\right).\frac{z}{c}.max\left(\frac{l_z}{l_c}.\frac{l_z}{r_c}.\frac{r_z}{l_c}.\frac{r_z}{r_c}\right)\right) \tag{6}$$

## 3. Fuzzy logistic regression model

The fuzzy logistic regression model consists of the effect of explanatory variables on the response variable Y. the response variable Y and the parameters $(\beta_0.\beta_1 \ldots\ldots \beta_k)$ represent the triangular fuzzy number while the explanatory variables $(X_1.X_2.\ldots.X_K)$ represent crisp numbers. so, the response function in the fuzzy parametric logistic regression model is according to the following formula: [18].

$$\tilde{\pi}_i = p_r(\tilde{y}_i = 1/X = x) = \frac{e^{\tilde{\beta}_0+\tilde{\beta}_1 X_{i1}+\cdots+\tilde{\beta}_k X_k}}{1 + e^{\tilde{\beta}_0+\tilde{\beta}_1 X_{i1}+\cdots+\tilde{\beta}_k X_k}} \tag{7}$$

Where i = 1, 2, 3 . . . n represents the number of observations and j = 1, 2, 3 . . . p represents the number of explanatory variables, and the matrix of explanatory variables represents X with dimensions (n*p), $\tilde{y}_i = (\tilde{y}; l_{\tilde{y}}.r_{\tilde{y}})$ the fuzzy triangular number dependent variable, $\tilde{\pi}_i = (\tilde{\pi}_i; l_{\tilde{\pi}_i}.r_{\tilde{\pi}_i})$ The fuzzy logistic regression function, $\underline{\tilde{\beta}} = (\tilde{\beta}_1.\tilde{\beta}_2.\ldots.\tilde{\beta}_p)$ The vector of undefined parameters i.e $\tilde{\beta}_j = (\tilde{b}_j; l_{\tilde{b}_j}.r_{\tilde{b}_j})$.

By applying the logit transformation, the curvature of the logistic function can be eliminated to stabilize the estimation process. if they exist. on the properties of the parameter estimates. The researcher Berkson was able in 1944 to find a logarithmic relationship to transform the response function $\tilde{\pi}_i$ into a linear

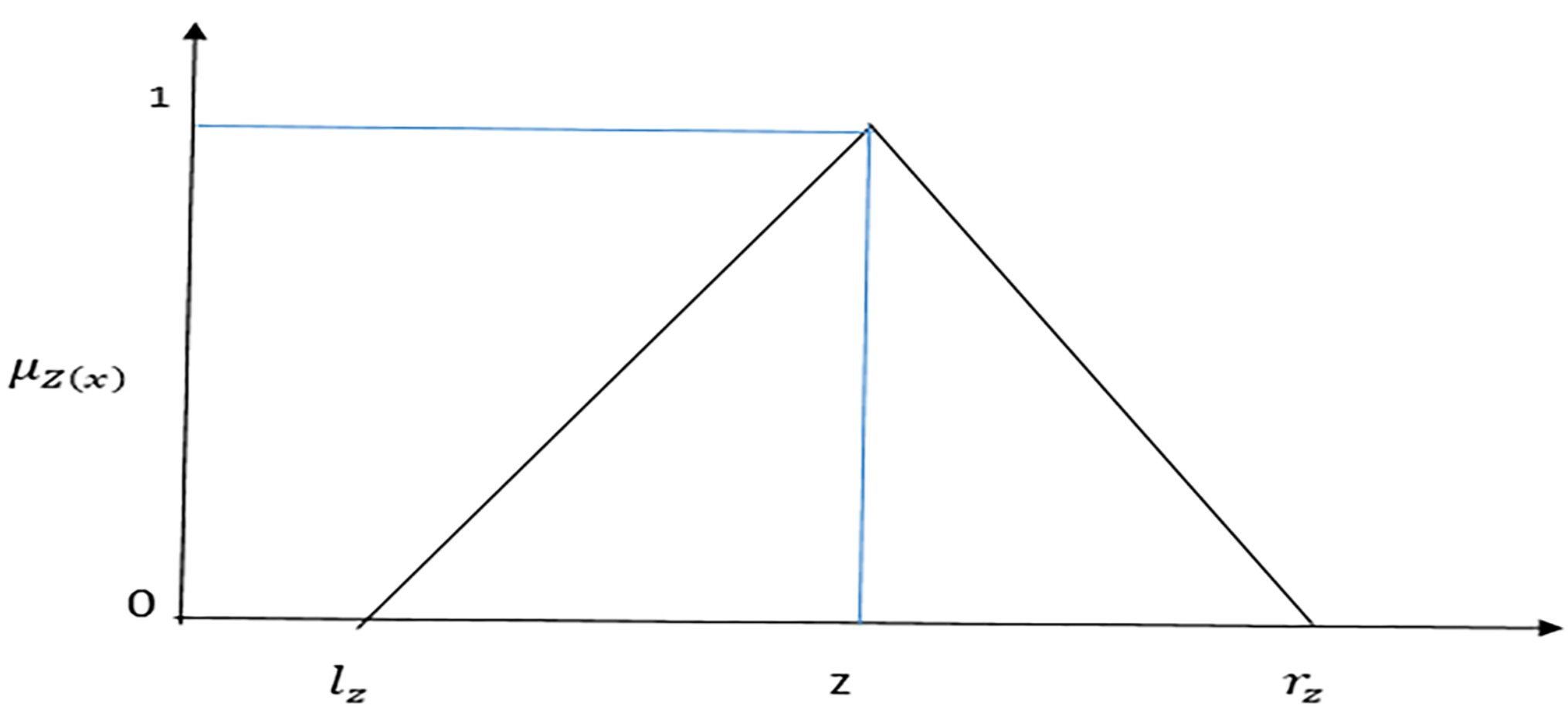

Fig. 1. Fuzzy triangular number [2].

function. Which is in the following form: [18].

$$logit\ (\tilde{\pi}_i) = ln\left(\frac{\tilde{\pi}_i}{1-\tilde{\pi}_i}\right) = X\underline{\tilde{\beta}} \qquad (8)$$

## 4. Estimating parameters of the fuzzy logistic regression model

### 4.1. Fuzzy maximum likelihood estimators

The goal of fuzzy maximum likelihood estimators is to find a set of estimated values for $\tilde{\boldsymbol{\beta}}$ (estimate the parameters that maximize the likelihood function). Fuzzy maximum likelihood estimators are used in methods for estimating the parameters of a logistic regression model in the case of a multicollinearity problem [15].

To get the maximum likelihood function for the fuzzy logistic regression model using the function

$$L\left(\tilde{\beta}.X\right) = \prod_{i=1}^{n} \tilde{\pi}_i^{\tilde{y}_i}(1-\tilde{\pi}_i)^{1-\tilde{y}_i} \qquad (9)$$

And taking the logarithm to the likelihood y function.

$$log L\left(\tilde{\beta}.X\right) = \sum_{i=1}^{n} log\,(1-\tilde{\pi}_i) + \left[\sum_{i=1}^{n} \tilde{y}_i X \underline{\tilde{\beta}})\right]$$

The solution cannot be found to obtain the parameters when finding the first derivative of the logarithm of the maximum likelihood function and setting it equal to zero. In such a case, the equations must be solved using numerical methods. The most common is the Newton-Raphson algorithm. Which is given by the equation:

$$\underline{\hat{\tilde{\beta}}}^{(t+1)}_{FMLE} = \underline{\tilde{\beta}}^{t} - I(\tilde{\beta})^{-1(t)} U\left(\underline{\tilde{\beta}}\right)^{(t)} \qquad (10)$$

Where $\underline{\tilde{\beta}}^{t}$ vector of fuzzy parameters in iteration t, has dimensions (P × 1), $U(\tilde{\beta})$ is the first derivative of the logarithm of the maximum likelihood function given by

$$U\left(\tilde{\beta}\right) = \acute{X}\left(y_i - \tilde{\pi}_i\right) \qquad (11)$$

$I(\tilde{\beta})^{-1}$ The inverse of the information matrix $I(\tilde{\beta})$ can be obtained from the second derivative of the logarithm of the maximum likelihood function of the fuzzy logistic regression model.

$$I\left(\tilde{\boldsymbol{\beta}}\right) = -\acute{X} diag \tilde{\boldsymbol{\pi}}_i \left(1-\tilde{\boldsymbol{\pi}}_i\right) X \qquad (12)$$

The Newton-Raphson method is iteratively applied until convergence is achieved to estimate the fuzzy maximum likelihood parameters $\underline{\hat{\tilde{\boldsymbol{\beta}}}}_{\boldsymbol{FMLE}}$ FMLE until the difference between two iterations in estimating the parameters is zero or close to zero.

### 4.2. Fuzzy-liu type estimators

These estimators are among the estimators that deal with multicollinearity of deterministic explanatory variables in the fuzzy logistic regression model. when the fuzzy maximum likelihood estimators are ineffective. which leads to the inflation of variances and the provision of misleading information.

A key feature of this method is its combination of ridge and Liu estimators to mitigate multicollinearity.

The ridge estimators for the logistic regression model were proposed by [14], The ridge estimators for the fuzzy logistic regression model **FLRE** are as follows:

$$\underline{\hat{\tilde{\beta}}}_{FLRE} = \left(X'\hat{W}X + KI\right)^{-1} X'\hat{W}X\underline{\hat{\tilde{\beta}}}_{FMLE} \qquad (13)$$

When the bias parameter K > 0, and the weight matrix $\hat{W}$ is a diagonal matrix whose main diagonal elements are estimated as $\tilde{\pi}_i(1-\tilde{\pi}_i)$.

These estimators extend the ridge approach originally proposed by Hoerl and Kennard (1970), and they have mean square error (MSE) according to the equation [10].

$$MSE\left(\underline{\hat{\tilde{\beta}}}_{FLRE}\right) = \left(X'\hat{W}X + KI\right)^{-1}\left(X'\hat{W}X\right)\left(X'\hat{W}X + KI\right)^{-1} \qquad (14)$$

The Liu logistic estimator (LLE) is an alternative estimator to the logistic ridge estimator to solve the multicollinearity problem, Therefore. The Liu estimator for the fuzzy logistic regression model (FLLE) is according to the equation [8]:

$$\underline{\hat{\tilde{\beta}}}_{FLLE} = \left(X'\hat{W}X + I\right)^{-1}\left(X'\hat{W}X + dI\right)\underline{\hat{\tilde{\beta}}}_{FMLE} \qquad (15)$$

When the bias parameter 0 < d < 1, and the mean square error (MSE) for the parameter estimators $\underline{\hat{\tilde{\beta}}}_{FLLE}$ is given by:

$$MSE\left(\underline{\hat{\tilde{\beta}}}_{FLLE}\right) = \left(X'\hat{W}X + I\right)^{-1}\left(X'\hat{W}X + dI\right)\left(X'\hat{W}X\right)^{-1} \times \left(X'\hat{W}X + dI\right)\left(X'\hat{W}X + I\right)^{-1}$$

As an alternative to the Ridge and Liu estimators, recent multicollinearity research has suggested estimators with two biases. By combining the two methods, these estimators hope to encourage the use of more appropriate estimators. A novel estimator based on the biases k and d was presented in this context by [9]. In order to solve the multicollinearity issue and lower variance, [3] has more recently adapted the Liu estimator into the Liu-type logistic estimator (LLTE) for binary logistic regression models, stabilizing the estimates in the process. As a result, the Liu-type estimators for the fuzzy logistic regression model (FLLTE) are as follows.

$$\underline{\widehat{\tilde{\beta}}}_{FLLTE} = \left(X'\hat{W}X + IK\right)^{-1}(X'\hat{W}X + dI)\underline{\widehat{\tilde{\beta}}}_{FMLE} \tag{16}$$

In addition, these parameter estimators $\underline{\widehat{\tilde{\beta}}}_{FLLTE}$ have a mean square error (MSE) [10].

$$MSE\left(\underline{\widehat{\tilde{\beta}}}_{FLLTE}\right) = \left(X'\hat{W}X + KI\right)^{-1}(X'\hat{W}X + dI) \times \left(X'\hat{W}X\right)^{-1}(X'\hat{W}X + dI)\left(X'\hat{W}X + KI\right)^{-1}$$

Furthermore, [8] proposed another biased estimator with two bias parameters when the two bias parameters are combined simultaneously so we will express this estimator (two-parameter Liu-type estimators for fuzzy logistic regression model (FLLTPE)) according to the following formula:

$$\underline{\widehat{\tilde{\beta}}}_{FLLTPE} = \left(X'\hat{W}X + IK\right)^{-1}(X'\hat{W}X + KdI)\underline{\widehat{\tilde{\beta}}}_{FMLE} \tag{17}$$

The mean square error of the fuzzy logistic Liu-type estimators is as follows: [10].

$$MSE\left(\underline{\widehat{\tilde{\beta}}}_{FLLPE}\right) = \left(X'\hat{W}X + KI\right)^{-1}(X'\hat{W}X + KdI) \times \left(X'\hat{W}X\right)^{-1}(X'\hat{W}X + KdI)\left(X'\hat{W}X + KI\right)^{-1}$$

### *4.3. Shrinkage parameters for fuzzy logistic Liu-type estimators*

The shrinkage parameters of the fuzzy logistic Liu-type estimators are calculated according to the following formulas: [17].

$$d = median\left[\frac{\lambda_j \alpha_j^2}{2\left(1 + \lambda_j \alpha_j^2\right)}\right] \tag{18}$$

$$d = min\left[\frac{\lambda_j \alpha_j^2}{2\left(1 + \lambda_j \alpha_j^2\right)}\right] \tag{19}$$

$$d = max\left[\frac{\lambda_j \alpha_j^2}{2\left(1 + \lambda_j \alpha_j^2\right)}\right] \tag{20}$$

When $\alpha_j = \gamma'\tilde{\beta}$ where $\gamma$ is a matrix whose columns represent the characteristic vectors of the matrix $X'\hat{W}X$ and $\lambda_j$ represents the characteristic roots of the matrix $X'\hat{W}X$ such that the condition $X'\hat{W}X = \gamma'\Lambda\gamma$ is satisfied where $\Lambda$ represents a diagonal matrix $\Lambda = daig(\lambda_1.\lambda_2.\ldots.\lambda_p)$ and j is from the elements to $0 < d < 1$ in addition to that $\lambda_j > 0$ and $\alpha_j^2 > 0$.

In addition to that. There are many formulas to calculate the shrinkage parameter d and the shrinkage parameter K is calculated depending on the shrinkage parameter d according to the following formula:

$$K = \frac{\lambda_j}{\lambda_j \alpha_j^2 (1 - d) - d} \tag{21}$$

## 5. Simulation study

In this section, based on the fuzzy logistic regression model, the estimators of the Liu type will be evaluated by comparing the estimators based on Monte Carlo simulation. Therefore, the stages of the simulation experiment will be described, taking into account the most important factors that affect the construction of the fuzzy logistic regression model by writing a program in the R language, which is one of the programs used in statistical programming and graphics, through the following steps.

1- The initial default values of the fuzzy parameters in the fuzzy logistic regression model are determined as shown in Table 1.

2- The crisp explanatory variables that contain the multicollinearity problem are generated by the following formula:

$$X_{ij} = \left(1 - \rho^2\right)^{0.5} U_{ij} + \rho U_{i(j+1)} \quad i = 1.2.3.\ldots.n \quad j = 1.2.3.\ldots.p \tag{22}$$

**Table 1.** Default values for fuzzy initial parameters.

| Model | $(\tilde{b}_0, l_{\tilde{b}_0}, r_{\tilde{b}_0})$ | $(\tilde{b}_1, l_{\tilde{b}_1}, r_{\tilde{b}_1})$ | $(\tilde{b}_2, l_{\tilde{b}_2}, r_{\tilde{b}_2})$ | $(\tilde{b}_3, l_{\tilde{b}_3}, r_{\tilde{b}_3})$ | $(\tilde{b}_4, l_{\tilde{b}_4}, r_{\tilde{b}_4})$ | $(\tilde{b}_5, l_{\tilde{b}_5}, r_{\tilde{b}_5})$ | $(\tilde{b}_6, l_{\tilde{b}_6}, r_{\tilde{b}_6})$ |
|---|---|---|---|---|---|---|---|
| 1 | (0.6,0.55,0.65) | (0.4,0.36,0.47) | (0.24,0.2,0.3) | (0.7,0.66,0.76) | (0.55,0.5,0.62) | (0.8,0.77,0.83) | (0.44,0.40,0.50) |
| 2 | (0.6,0.55,0.65) | (0.4,0.36,0.47) | (0.24,0.2,0.3) | (0.7,0.66,0.76) | (0.55,0.5,0.62) | | |
| 3 | (0.6,0.55,0.65) | (0.4,0.36,0.47) | (0.24,0.2,0.3) | (0.7,0.66,0.76) | | | |

**Table 2.** Represents different impact factors.

| The Effect | Values |
|---|---|
| Number of Explanatory Variables k | 3, 4, 7 |
| Sample Size n | 25, 50, 100, 150 |
| Correlation coefficient $\rho$ | 0.70, 0.90, 0.99 |

When $\rho$ the correlation coefficient between the crisp explanatory variables. $U_{ij}$ Represents random numbers that follow the standard normal distribution $U_{ij} \sim (0.1)$.

3- Generating values of the fuzzy random error variable in a fuzzy logistic regression model according to the Bernoulli distribution

$$e_i \sim bernoulli\,(0,\ 0.7)$$

$$l_{e_i} \sim bernoulli\,(0,\ 0.5)$$

$$r_{e_i} \sim bernoulli\,(0,\ 0.8)$$

4- The fuzzy dependent variable $\tilde{y}_i$ is calculated according to the inverse transformation generation method based on the fuzzy logistic regression function and the random error term according to the following formula:

$$\tilde{y}_i = logit\ \ (\tilde{\pi}_i) + e_i \tag{23}$$

5- The data generation process was repeated for different sample sizes, numbers of explanatory variables, and correlation coefficients, over 100 iterations according to the following Table 2.

6- The simulation used bias parameter d as specified in method 1, where the median shrinkage value was applied.

7- The comparison among the fuzzy logistic regression models was conducted using the Mean Square Error (MSE) and Goodness-of-Fit criteria, which depend on calculating the Euclidean-distances according to the following formula: [6].

$$d_i\left(\tilde{y}_i.\widehat{\tilde{y}}_i\right) = \sqrt[2]{\frac{1}{3}(y-\hat{y})^3 + (l - l_{\hat{y}})^2 + (r_y - r_{\hat{y}})^2}$$

Then the mean square error and goodness of fit are calculated according to the following formulas:

$$MSE\left(\tilde{y}_i.\widehat{\tilde{y}}_i\right) = \frac{1}{1000}\sum d_i\left(\tilde{y}_i.\widehat{\tilde{y}}_i\right)$$

$$S\left(\tilde{y}_i.\widehat{\tilde{y}}_i\right) = \frac{1}{1000}\sum \frac{1}{1 + d_i\left(\tilde{y}_i.\widehat{\tilde{y}}_i\right)}$$

The Table 3 represents the comparison criteria of the mean square error and goodness of fit for the fuzzy logistic regression model estimation methods at different sample sizes and different correlation coefficients for the explanatory variables. We note through the comparison criteria that the efficiency of the FLLTPE and FLLTE methods is superior to the rest of the methods in all sample sizes and for all correlation coefficients, while the effectiveness of the fuzzy maximum likelihood estimators appeared less efficient.

From the Table 4, when reducing the number of variables to 5 explanatory variables in the model 2. It is observed that a slight change in the comparison criteria, as the FLLTPE are still in the lead, more effective and efficient, followed by the FLLTE estimators compared to the model 1 while we see an increase in the mean square error and goodness of fit criteria for the FLLE and FMLE methods.

From Table 5, we note the superiority of the FLLTPE method and then the FLLTE method based on the standard of the mean square error and goodness of fit for all sample sizes and correlation coefficients, but the difference between the comparison standards for all estimation methods was less than in the models 1 and 2.

**Table 3.** Mean square error and goodness of fit for fuzzy logistic regression model when number of explanatory variables 7 (model 1).

| methods | | *FMLE* | | *FLRE* | | *FLLE* | | *FLLTE* | | *FLLTPE* | |
|---|---|---|---|---|---|---|---|---|---|---|---|
| n | $\rho$ | $S(\tilde{y}_i, \widehat{\tilde{y}}_i)$ | *MSE* | $S(\tilde{y}_i, \widehat{\tilde{y}}_i)$ | *MSE* | $S(\tilde{y}_i, \widehat{\tilde{y}}_i)$ | *MSE* | $S(\tilde{y}_i, \widehat{\tilde{y}}_i)$ | *MSE* | $S(\tilde{y}_i, \widehat{\tilde{y}}_i)$ | *MSE* |
| n = 25 | $\rho = 0.70$ | 1.7567 | 2.502 | 1.3123 | 2.0576 | 0.9345 | 1.6798 | 0.3456 | 1.0909 | 0.2546 | 1.0332 |
| | $\rho = 0.90$ | 1.8345 | 2.5798 | 1.4234 | 2.1687 | 1.0234 | 1.7687 | 0.4234 | 1.1687 | 0.3421 | 1.0874 |
| | $\rho = 0.99$ | 1.9456 | 2.6909 | 1.5345 | 2.2798 | 1.1345 | 1.8798 | 0.5345 | 1.2798 | 0.4123 | 1.1576 |
| n = 50 | $\rho = 0.70$ | 1.856 | 2.602 | 1.4678 | 2.2131 | 1.0678 | 1.8131 | 0.4789 | 1.2242 | 0.3567 | 1.1021 |
| | $\rho = 0.90$ | 1.9456 | 2.6909 | 1.5678 | 2.3131 | 1.1456 | 1.8978 | 0.5678 | 1.3131 | 0.4789 | 1.2242 |
| | $\rho = 0.99$ | 2.0234 | 2.7687 | 1.6456 | 2.3909 | 1.2345 | 1.9798 | 0.6456 | 1.3909 | 0.5567 | 1.3020 |
| $n = 100$ | $\rho = 0.70$ | 1.9456 | 2.6909 | 1.5678 | 2.3131 | 1.1456 | 1.8909 | 0.5678 | 1.3131 | 0.4678 | 1.2131 |
| | $\rho = 0.90$ | 2.0345 | 2.7798 | 1.6567 | 2.4020 | 1.2234 | 1.9798 | 0.6567 | 1.4020 | 0.5789 | 1.3242 |
| | $\rho = 0.99$ | 2.1123 | 2.8576 | 1.7345 | 2.4798 | 1.3123 | 2.0678 | 0.7345 | 1.4798 | 0.6678 | 1.4131 |
| n = 150 | $\rho = 0.70$ | 2.0234 | 2.7687 | 1.7345 | 2.4242 | 1.2234 | 1.9798 | 0.6789 | 1.4242 | 0.5789 | 1.3242 |
| | $\rho = 0.90$ | 2.1012 | 2.8465 | 1.7567 | 2.5020 | 1.3012 | 2.0678 | 0.7567 | 1.5020 | 0.6789 | 1.4242 |
| | $\rho = 0.99$ | 2.2011 | 2.9464 | 1.8456 | 2.5909 | 1.4011 | 2.1467 | 0.8456 | 1.5909 | 0.7890 | 1.5343 |

**Table 4.** Mean square error and goodness of fit for fuzzy logistic regression model when number of explanatory variables 5 (model 2).

| methods | | *FMLE* | | *FLRE* | | *FLLE* | | *FLLTE* | | *FLLTPE* | |
|---|---|---|---|---|---|---|---|---|---|---|---|
| n | $\rho$ | $S(\tilde{y}_i, \widehat{\tilde{y}}_i)$ | *MSE* | $S(\tilde{y}_i, \widehat{\tilde{y}}_i)$ | *MSE* | $S(\tilde{y}_i, \widehat{\tilde{y}}_i)$ | *MSE* | $S(\tilde{y}_i, \widehat{\tilde{y}}_i)$ | *MSE* | $S(\tilde{y}_i, \widehat{\tilde{y}}_i)$ | *MSE* |
| n = 25 | $\rho = 0.70$ | 1.5678 | 2.3131 | 1.3456 | 2.0909 | 1.0345 | 1.7798 | 0.6578 | 1.3131 | 0.4512 | 1.1965 |
| | $\rho = 0.90$ | 1.6123 | 2.3577 | 1.3890 | 2.1343 | 1.0789 | 1.8242 | 0.6123 | 1.3577 | 0.4876 | 1.2329 |
| | $\rho = 0.99$ | 1.6789 | 2.4242 | 1.4345 | 2.1798 | 1.1234 | 1.8687 | 0.6456 | 1.3910 | 0.5234 | 1.2687 |
| n = 50 | $\rho = 0.70$ | 1.4789 | 2.2242 | 1.2567 | 2.0020 | 0.9456 | 1.6909 | 0.4789 | 1.2242 | 0.3789 | 1.1242 |
| | $\rho = 0.90$ | 1.5345 | 2.2798 | 1.3234 | 2.0687 | 1.0234 | 1.7798 | 0.5345 | 1.2798 | 0.4234 | 1.1687 |
| | $\rho = 0.99$ | 1.6012 | 2.3465 | 1.3789 | 2.1242 | 1.0789 | 1.8242 | 0.5678 | 1.3131 | 0.4567 | 1.2020 |
| $n = 100$ | $\rho = 0.70$ | 1.4234 | 2.1687 | 1.2012 | 1.9465 | 0.8901 | 1.6354 | 0.4234 | 1.1687 | 0.3123 | 1.0576 |
| | $\rho = 0.90$ | 1.4789 | 2.2242 | 1.2567 | 2.0020 | 0.9345 | 1.6798 | 0.4678 | 1.2131 | 0.3456 | 1.0909 |
| | $\rho = 0.99$ | 1.5345 | 2.2798 | 1.3123 | 2.0576 | 0.9789 | 1.7242 | 0.5012 | 1.2465 | 0.3890 | 1.1343 |
| n = 150 | $\rho = 0.70$ | 1.3890 | 2.1343 | 1.1678 | 1.9124 | 0.8567 | 1.6021 | 0.3890 | 1.1343 | 0.2890 | 1.0343 |
| | $\rho = 0.90$ | 1.4234 | 2.1687 | 1.2012 | 1.9465 | 0.8901 | 1.6354 | 0.4234 | 1.1687 | 0.3123 | 1.0576 |
| | $\rho = 0.99$ | 1.4789 | 2.2242 | 1.2567 | 2.0020 | 0.9345 | 1.6798 | 0.4678 | 1.2131 | 0.3456 | 1.0909 |

**Table 5.** Mean square error and goodness of fit for fuzzy logistic regression model when number of explanatory variables 3 (model 3).

| methods | | *FMLE* | | *FLRE* | | *FLLE* | | *FLLTE* | | *FLLTPE* | |
|---|---|---|---|---|---|---|---|---|---|---|---|
| n | $\rho$ | $S(\tilde{y}_i, \widehat{\tilde{y}}_i)$ | *MSE* | $S(\tilde{y}_i, \widehat{\tilde{y}}_i)$ | *MSE* | $S(\tilde{y}_i, \widehat{\tilde{y}}_i)$ | *MSE* | $S(\tilde{y}_i, \widehat{\tilde{y}}_i)$ | *MSE* | $S(\tilde{y}_i, \widehat{\tilde{y}}_i)$ | *MSE* |
| n = 25 | $\rho = 0.70$ | 1.7365 | 2.5064 | 1.4852 | 2.2915 | 1.1231 | 2.0674 | 0.6622 | 1.4987 | 0.5265 | 1.3629 |
| | $\rho = 0.90$ | 1.8251 | 2.5919 | 1.5745 | 2.3758 | 1.2105 | 2.1539 | 0.7471 | 1.5835 | 0.6084 | 1.4442 |
| | $\rho = 0.99$ | 1.9262 | 2.7032 | 1.6756 | 2.4884 | 1.3097 | 2.2671 | 0.8497 | 1.6843 | 0.7091 | 1.5385 |
| n = 50 | $\rho = 0.70$ | 1.6862 | 2.4558 | 1.4381 | 2.2415 | 1.0781 | 2.0139 | 0.6164 | 1.4585 | 0.4794 | 1.3184 |
| | $\rho = 0.90$ | 1.7621 | 2.5351 | 1.5227 | 2.3220 | 1.1557 | 2.0954 | 0.6903 | 1.5339 | 0.5557 | 1.3925 |
| | $\rho = 0.99$ | 1.8695 | 2.6565 | 1.6355 | 2.4397 | 1.2542 | 2.2025 | 0.7881 | 1.6410 | 0.6532 | 1.4962 |
| $n = 100$ | $\rho = 0.70$ | 1.6082 | 2.3962 | 1.3718 | 2.1734 | 1.0171 | 1.9650 | 0.5565 | 1.3987 | 0.4216 | 1.2571 |
| | $\rho = 0.90$ | 1.6871 | 2.4751 | 1.4503 | 2.2544 | 1.0978 | 2.0479 | 0.6369 | 1.4751 | 0.4942 | 1.3348 |
| | $\rho = 0.99$ | 1.7961 | 2.5951 | 1.5598 | 2.3713 | 1.1954 | 2.1564 | 0.7310 | 1.5792 | 0.5824 | 1.4300 |
| n = 150 | $\rho = 0.70$ | 1.5536 | 2.3482 | 1.3197 | 2.1260 | 0.9874 | 1.9319 | 0.5207 | 1.3482 | 0.3785 | 1.2049 |
| | $\rho = 0.90$ | 1.6199 | 2.4269 | 1.3862 | 2.2044 | 1.0492 | 2.0133 | 0.5895 | 1.4137 | 0.4463 | 1.2710 |
| | $\rho = 0.99$ | 1.7365 | 2.5064 | 1.4852 | 2.2915 | 1.1231 | 2.0674 | 0.6622 | 1.4987 | 0.5265 | 1.3629 |

## 6. Application of real data

To assess the performance of the proposed fuzzy logistic regression estimators, real clinical data were collected from Al-Zahraa Hospital in Al-Kut city, comprising a sample of 35 patients diagnosed with varying degrees of kidney failure. The dataset was compiled based on expert evaluations from a team of medical professionals who confirmed that kidney failure progression is inherently imprecise and best represented using fuzzy linguistic terms.

The response variable renal failure rate was expressed as a fuzzy triangular number according to expert classification, as shown in Table 6:

**Table 6.** The fuzzy triangular number (fuzzy response variable).

| Fuzzy Terms | Values |
|---|---|
| Low | (0, 0.25, 0.44) |
| Medium | (035, 0.55, 0.74) |
| High | (0.65, 0.85, 1) |

The explanatory variables include:

X1: Blood sugar level (indicator of diabetes), normally ranging between 90–120 mg/dL.
X2: Hemoglobin level (Hb), measured in g/dL.
X3: Patient age in years.

These variables are known to contribute significantly to the incidence of kidney failure and are presented in Table 7.

To examine the presence of multicollinearity among the explanatory variables, eigenvalues, condition numbers, and variance proportions were calculated as shown in Table 8.

From Table 8, the condition number corresponding to variable X3 was 45.5758 exceeding the critical value of 30-indicating a serious multicollinearity problem. Furthermore, the variance proportions revealed that variables X1 and X3 share a high variance in the same principal component direction, further confirming multicollinearity.

The fuzzy logistic regression model was then applied using various estimation techniques, and the results are reported in Table 9.

The findings demonstrate that the Fuzzy Logistic Liu-Type Parameter Estimator (FLLTPE) and the Fuzzy Logistic Liu-Type Estimator (FLLTE) achieved the best performance in terms of minimizing the Mean Square Error (MSE) and maximizing the

**Table 7.** The incidence of kidney failure and explanatory variables.

| N | Renal failure rate | $X_1$ | $X_2$ | $X_3$ |
|---|---|---|---|---|
| 1 | Low | 123 | 9 | 60 |
| 2 | High | 178 | 7.6 | 67 |
| 3 | Medium | 140 | 10 | 70 |
| 4 | High | 167 | 9 | 100 |
| 5 | High | 155 | 6 | 89 |
| 6 | Low | 133 | 9.6 | 77 |
| 7 | Medium | 145 | 11.2 | 80 |
| 8 | High | 178 | 10.1 | 76 |
| 9 | High | 180 | 8.4 | 60 |
| 10 | Medium | 144 | 11.5 | 59 |
| 11 | Medium | 157 | 8.2 | 76 |
| 12 | High | 145 | 13 | 89 |
| 13 | Medium | 167 | 8.1 | 79 |
| 14 | High | 180 | 10 | 90 |
| 15 | Medium | 135 | 9.8 | 77 |
| 16 | High | 155 | 11.2 | 90 |
| 17 | High | 169 | 8.9 | 88 |
| 18 | Medium | 155 | 8.4 | 87 |
| 19 | Low | 123 | 12 | 78 |
| 20 | Low | 144 | 10.3 | 90 |
| 21 | Medium | 165 | 12.2 | 83 |
| 22 | Low | 163 | 9.6 | 79 |
| 23 | Medium | 134 | 8.7 | 92 |
| 24 | Low | 140 | 12 | 95 |
| 25 | High | 120 | 10.4 | 72 |
| 26 | High | 137 | 11.8 | 76 |
| 27 | Low | 155 | 9.2 | 89 |
| 28 | Medium | 150 | 11.9 | 95 |
| 29 | Low | 157 | 10.3 | 77 |
| 30 | Medium | 145 | 9.4 | 94 |
| 31 | Low | 122 | 12 | 81 |
| 32 | Medium | 120 | 8 | 87 |
| 33 | Low | 130 | 9.7 | 79 |
| 34 | High | 156 | 11 | 100 |
| 35 | Low | 123 | 7.8 | 77 |

**Table 8.** The conditional number and the variance ratio.

| Independent variable | Eigenvalue | Condition number | variance proportion | | | |
|---|---|---|---|---|---|---|
| $X_0$ | 5.99381 | 1.00000 | 0.049 | 0.258 | 0.006 | 0.263 |
| $X_1$ | 0.09055 | 8.79410 | 0.029 | 0.531 | 0.073 | 0.214 |
| $X_2$ | 0.03431 | 15.1870 | 0.321 | 0.021 | 0.336 | 0.135 |
| $X_3$ | 0.00207 | 45.5758 | 0.028 | 0.112 | 0.214 | 0.073 |

Goodness of Fit. In contrast, the Fuzzy Maximum Likelihood Estimator (FMLE) recorded the highest MSE and the poorest model fit, confirming its weakness under multicollinearity conditions.

These results support the advantage of Liu-type estimators in fuzzy modeling when multicollinearity exists among the explanatory variables, particularly in sensitive domains like medical diagnostics.

### Discussion of the results

Together, the simulation and real-data results show how multicollinearity significantly affects fuzzy logistic regression models and highlight how well Liu-type estimators work to mitigate this problem.

**Table 9.** Represent fuzzy parameter estimation, mean square error and goodness of fit for the fuzzy logistic regression model.

| methods | | *FMLE* | *FLRE* | *FLLE* | *FLLTE* | *FLLTPE* |
|---|---|---|---|---|---|---|
| $\widehat{\tilde{\beta}}_0$ | $\widehat{\tilde{\beta}}_0$ | 3.263152 | 2.03947 | 2.266078 | 1.812862 | 1.586255 |
| | $l_{\widehat{\tilde{\beta}}_0}$ | 3.14581 | 1.966131 | 2.18459 | 1.747672 | 1.529213 |
| | $r_{\widehat{\tilde{\beta}}_0}$ | 3.934974 | 2.459359 | 2.732621 | 2.186097 | 1.912835 |
| $\widehat{\tilde{\beta}}_1$ | $\widehat{\tilde{\beta}}_1$ | 4.543368 | 2.839605 | 3.155117 | 2.524094 | 2.208582 |
| | $l_{\widehat{\tilde{\beta}}_1}$ | 4.512826 | 2.820516 | 3.133907 | 2.507126 | 2.193735 |
| | $r_{\widehat{\tilde{\beta}}_1}$ | 5.961496 | 3.725935 | 4.139928 | 3.311942 | 2.89795 |
| $\widehat{\tilde{\beta}}_2$ | $\widehat{\tilde{\beta}}_2$ | 0.464881 | 0.290551 | 0.322834 | 0.258267 | 0.225984 |
| | $l_{\widehat{\tilde{\beta}}_2}$ | 0.185194 | 0.115746 | 0.128607 | 0.102886 | 0.090025 |
| | $r_{\widehat{\tilde{\beta}}_2}$ | 0.471666 | 0.294791 | 0.327546 | 0.262037 | 0.229282 |
| $\widehat{\tilde{\beta}}_3$ | $\widehat{\tilde{\beta}}_3$ | 0.683421 | 0.427138 | 0.474598 | 0.379678 | 0.332219 |
| | $l_{\widehat{\tilde{\beta}}_3}$ | 0.732084 | 0.457553 | 0.508392 | 0.406714 | 0.355874 |
| | $r_{\widehat{\tilde{\beta}}_3}$ | 1. 84216 | 0.72345 | 0. 84216 | 0.674533 | 0.624579 |
| MS | | 1.8073 | 1.4152 | 1.2882 | 0.5672 | 0.5342 |
| $S(\tilde{y}_i.\widehat{\tilde{y}}_i)$ | | 0.54219 | 0.48456 | 0.38646 | 0.17016 | 0.16026 |

The Fuzzy Logistic Liu-Type Parameter Estimator (FLLTPE) consistently performed better than the other five estimation methods that were evaluated: FMLE, FLRE, FLLE, FLLTE, and FLLTPE. This was demonstrated in both the real medical dataset on kidney failure and the simulated datasets, where FLLTPE yielded the highest Goodness of Fit (GOF) and the smallest Mean Square Error (MSE).

Despite its theoretical soundness, the conventional fuzzy maximum likelihood estimator (FMLE) performed the worst in multicollinearity scenarios, as evidenced by higher MSE values and a worse model fit. This demonstrates its limited robustness in fuzzy environments and its sensitivity to predictor correlation.

On the other hand, FLLTPE and FLLTE both showed accuracy and stability across a range of models and sample sizes. Their better performance can be explained by the bias-variance trade-off mechanisms that are built into their design to account for collinearity.

These results validate the suitability of Liu-type estimators, particularly the parameter-adjusted versions, for fuzzy regression models in real-world settings with multicollinearity and uncertain inputs.

## 7. Conclusions

The following is a summary of the main findings from this study:

1. The Fuzzy Logistic Liu-Type Parameter Estimator (FLLTPE) and the Fuzzy Logistic Liu-Type Estimator (FLLTE) consistently outperformed other estimators based on the Mean Square Error (MSE) and Goodness of Fit (GOF) criteria. In a variety of scenarios, these techniques produced the best model fit and the lowest MSE.

2. For all tested models, increasing the sample size from 25 to 150 resulted in appreciable improvements in GOF and decreases in MSE. This pattern attests to the statistical models' increased accuracy and resilience when used on bigger datasets.
3. While the GOF shows how well the model fits the observed data, the MSE is a trustworthy measure of average prediction error. MSE values were consistently higher than GOF values across all estimation techniques and sample sizes, suggesting a typical pattern in model evaluation.
4. Among the explanatory variables, a multicollinearity issue was found, especially between variables X1 (diabetes) and X3 (patient age). High condition numbers (above 30) and the variance proportion matrix analysis, which indicated strong linear dependency, were used to identify this problem.
5. The fuzzy parameters across all Liu-type estimators support medical observations by showing that diabetes significantly contributes to kidney failure.
6. Future studies could look into different fuzzy estimation methods or use the current approach on more complicated and sizable real-world datasets, especially in the biomedical field.

## Funding

The authors declare that no funds, grants, or other support were received during the preparation of this manuscript.

## Conflicts of interest

The writers affirm that they have no competing interests.

## Acknowledgment

The authors thank their respective institutions for their assistance.